\documentclass[review]{elsarticle}

\usepackage{lineno,hyperref,multirow}
\journal{Materials Research Bulletin}

\def\unitangstrom{\,\textrm{\AA}}
\def\unitseebeck{\, {\rm \mu V/K}}
\def\unitev{\,{\rm eV}}

\def\cmo{\rm CaMnO_3}
\def\bcmo{\rm Bi_{0.03}Ca_{0.97}MnO_3}

\newcommand{\lp}{\left(}
\newcommand{\rp}{\right)}
\renewcommand{\epsilon}{\varepsilon}

\begin{document}

\begin{frontmatter}

\title{Investigation of electron and phonon transport in Bi-doped
  CaMnO$_3$ for thermoelectric applications}
  

\author[lipi]{Edi~Suprayoga}
\ead[e-mail]{edis008@lipi.go.id}

\author[lipi]{Witha~B.~K.~Putri}
\author[snru1]{Kunchit~Singsoog}
\author[snru1,snru2]{Supasit~Paengson}
\author[lipi]{Muhammad~Y.~Hanna}
\author[lipi]{Ahmad~R.~T.~Nugraha}
\author[ui]{Dicky~R.~Munazat}
\author[ui]{Budhy~Kurniawan}
\author[ub]{Muhammad~Nurhuda}
\author[snru1,snru2]{Tosawat~Seetawan}
\ead[e-mail]{t\_seetawan@snru.ac.th}

\author[lipi,lux]{Eddwi~H.~Hasdeo}
\ead[e-mail]{eddw001@lipi.go.id}

\address[lipi]{Research Center for Physics, Indonesian Institute of
  Sciences (LIPI), Tangerang Selatan 15314, Indonesia}
\address[snru1]{Center of Excellence on Alternative Energy, Research and
  Development Institution, Sakon Nakhon Rajabhat University, Sakon
  Nakhon 47000, Thailand}
\address[snru2]{Program of Physics, Faculty of Science and Technology,
  Sakon Nakhon Rajabhat University, Sakon Nakhon 47000, Thailand}
\address[ui]{Department of Physics, Universitas Indonesia, Depok 16424, Indonesia}
\address[ub]{Department of Physics, Brawijaya University, Malang,
  65145, Indonesia}
\address[lux]{Physics and Materials Science Research Unit, University
  of Luxembourg, L-1511, Luxembourg}

\begin{abstract}
 The electron and phonon transports in $\cmo$ and in one of its Bi-doped counterparts, namely, $\bcmo$, are investigated using the thermoelectric transport measurements and first-principles calculations.  We find that antiferromagnetic insulator $\cmo$ breaks the Wiedemann--Franz law with the Lorenz number reaching four times that of ordinary metals at room temperature. Bismuth doping reduces both the electrical resistivity and Seebeck coefficient of $\cmo$; thus, it recovers the Wiedemann--Franz law behavior.  In addition, $\bcmo$ possesses a shorter phonon lifetime according to the transport measurements.  As a result, $\bcmo$ exhibits superior thermoelectric properties over pristine $\cmo$ owing to the lower thermal conductivity and electrical resistivity.
\end{abstract}


\end{frontmatter}

%\linenumbers

\section{Introduction}
Perovskite oxides have attracted extensive attention in the last
decades owing to their rich physical
properties~\cite{Oz2018, cong2018, Neetika2012}.
Strongly correlated electrons that originate from transition-metal atoms
cause various ground states, such as
antiferromagnetic (AFM) insulators~\cite{abrashev, Wiebe2001}, ferromagnetic
metals~\cite{Lone2019}, and high-$T_c$
superconductors~\cite{rao1990}.  These strongly correlated
perovskite oxides can, in particular, be applied to sensor technology by 
utilizing their piezoelectric and giant
magnetoresistance~\cite{zheng2010, Zeng1999} as well as in energy-harvesting
technology, e.g., in solar cells and thermoelectric (TE) devices
~\cite{baranovskiy, molinari2014, seetawan, paengson}.

TE materials are a class of materials that can convert heat into
electricity~\cite{goldsmid,zebarjani}.  Efficient TE materials are
thermally inert while simultaneously being electrically conducting.
These requirements provide researchers a challenge in designing TE
materials because \textbf{the moving electrons that constitute current in regular metals also carry heat}.  According to the Wiedemann--Franz (WF)
law, the ratio of electron thermal conductivity $\kappa_e$ to
electrical conductivity $\sigma$ is proportional to the 
temperature~\cite{wiedemann}. Thus, a possible method of improving the TE
efficiency is by breaking the WF law.  Breaking the WF law
might be realized, e.g., by introducing strongly correlated
Coulomb interactions~\cite{mahajan2013,principi2015,lavasani2019}.

Perovskite oxides are considered strongly correlated electronic
systems.  In this regard, we propose that one of the simplest
perovskite oxides, i.e., $\rm CaMnO_3$ (or ``CMO") with proper doping can
be a potential candidate for nontoxic and easily prepared TE
materials.  CMO is structurally stable even at temperatures above
$900^{\circ}\textrm{C}$; thus, it is suitable for high-temperature TE
materials~\cite{Lin}.  At its ground state, CMO behaves as
an insulator with an AFM order that originates from the Mott-like
correlation~\cite{molinari2014,spaldin}.  
\textbf{The strongly correlated interaction in this system is responsible for the gap opening and AFM ground states of CMO.}
At higher temperature, its
electrical conductivity increases, thus allowing better TE
performance.  
When it is doped with $\rm Bi$ into the $\rm Ca$ site, a 
double-exchange interaction between two different Mn ions, which is mediated by
oxygen, delocalizes electrons, which allows metal ferromagnetic ground states~\cite{anderson}.  We must note that whereas electronic
contribution to the transport properties has been well
understood, the phonon contribution to the thermal conductivity and
the role of Bi doping in the phonon transport in CMO remain unclear.

The present study aims to investigate the electron and phonon 
transport properties in CMO and Bi-doped CMO (abbreviated as ``BCMO") 
using a combination of TE transport measurements and first-principles calculations of the Seebeck coefficient, electrical conductivity, and thermal conductivity.
We choose $\bcmo$ as BCMO in the present study because of
our expectation that this particular compound provides the best TE
performance among various compositions of BCMO~\cite{paengson}.  The
Seebeck coefficients, electrical resistivities, and thermal
conductivities are measurable quantities in the experiments compared
with the first-principles calculations.  By calculating the electronic
structure and phonon energy of the materials (and comparing them with the
experimental results), we can possibly obtain essential quantities
that determine the transport properties of materials, such as the chemical
potential, electron lifetime, phonon lifetime, phonon Debye
temperature, and sound velocities.  As a supplement, we also
introduce Raman spectroscopy to understand the role of Bi doping in
the phonon transport.

This paper is organized as follows.  The experimental and theoretical
methods are presented in Sections~\ref{sec:exp} and ~\ref{sec:theo},
respectively.  The geometrical CMO structure is confirmed using X-ray
diffraction (XRD), as discussed in Section~\ref{sec:geo}. The electronic
transport properties of CMO and BCMO are presented in
Sections~\ref{sec:el1} and ~\ref{sec:el2}, respectively.  The phonon
contribution to thermal conductivity is presented in
Section~\ref{sec:ph}.  Finally, conclusions are given in
Section~\ref{sec:conclusion}.

\section{Experimental Methods}
\label{sec:exp}

To prepare for the characterization of the TE properties, we synthesized the CMO and BCMO materials using the solid-state reaction and hot-press methods.  The raw materials, including $\rm CaCO_3$ (from QReC, purity $\geq 99\%$), $\rm MnO_2$ (Fisher Scientific; purity $\geq 99.99\%$), and $\rm Bi_2O_3$ (QReC, purity $\geq 99.5\%$), were mixed using a planetary ball mill (RETSCH PM 400) in ethanol for $12$ h. The mixed powder was calcined at $1273$ K for $24$ h in the atmosphere and then hot-pressed at $1173$ K for an hour under $60$ MPa pressure in argon atmosphere. The hot-pressed pellets were annealed at $1473$ K for $36$ h in air.  We pulverized the pellets into powder for \textbf{crystal-structure analysis} using the XRD technique (using Shimadzu 6100) with CuK$\alpha$ radiation. The scan speed is $2^{\circ}$/min, whereas the scan range is within $2~\theta$ of $20^{\circ}$ -- $80^{\circ}$.

The Seebeck coefficient ($S$) and electrical resistivity ($\rho$) were
measured using the four-point probe method at a temperature range of $300$--$473$ K in the
atmosphere. The values of $S$ in this experiment were obtained from the formula 
$S = -\Delta V/\Delta T$, where $T$ is temperature and $V$ is the voltage. 
We obtain the values of $\rho$ from the formula $\rho=RA/l$, where R is the resistance, $A$ is the cross-sectional area of the material, and $l$ is the probe distance.  The thermal conductivity ($\kappa$) was measured using the steady-state method according to the relationship $\kappa=-\dot Ql/A\Delta T$, where $\dot Q$ is the heat rate.

\section{Theoretical Methods}
\label{sec:theo}

To obtain accurate geometry and electronic structures of the materials, we perform first-principles density-functional theory (DFT) calculations using the projected augmented wave and spin-polarized bases, as implemented in the \texttt{Quantum ESPRESSO} package~\cite{QE}.  To consider the exchange-correlation effects, we employ the revised generalized-gradient approximation proposed by Perdew, Burke, and Ernzerhof, which was implemented in the so-called PBEsol functional~\cite{PBE} and was appropriate for a densely packed solid surface.  At the ground state, CMO is an AFM
material~\cite{ling2003}, and it belongs to a \emph{Pnma} space group
containing $20$ atoms in the unit cell (see Fig.~\ref{xrd-cmo}). According to a previous study by Molinari
\emph{et al.}~\cite{molinari2014}, we
add Coloumb interaction potential $U=5\unitev$ on the DFT+U level to account for the localized
$\textrm{Mn}$ states.  The converged result is obtained using the $6 \times 6 \times 6$ \emph{k}-mesh sampling in the
Monkhorst--Pack scheme and kinetic-energy cutoff of $65$ Ry.  All
atoms are relaxed until all forces are smaller than $1\times10^{-5}$
Ry/a.u.  The density of states (DOS), which is defined as the number of states at given energy $E$, is obtained using the tetrahedron method through the formula $\mathrm{DOS}(E)=\sum^{~}_{i{\mathbf{k}}} \delta(E_{i\mathbf{k}} - E)$~\cite{tetrahedron}, where $E_{i\mathbf{k}}$ is the energy dispersion relationship in the $i$-th electronic-energy band.  We also calculate the phonon
frequencies and eigenvectors of CMO at the $\Gamma$ point (center of the Brillouin zone).
The force constant is obtained from the density-functional
perturbation theory (DFPT)+U scheme~\cite{dfpt}, whereas the phonon frequencies and
eigenvectors can be obtained by solving the dynamic matrix equation.

To understand the role of Bi doping in the electronic transport, we then calculate the electronic structure of BCMO by considering the $2 \times 2 \times 2$ supercell of CMO consisting of $32$ Ca atoms. We replace one Ca atom at the center of the supercell with one Bi atom to obtain 3.125\% Bi doping.  From the electronic
structures ($E_{i\mathbf{k}}$) of CMO and BCMO, we can calculate the Seebeck coefficient
($S$), electrical conductivity ($\sigma$), and electronic thermal
conductivity ($\kappa_e$) at the measurement temperature in the linear
Boltzmann transport theory and the relaxation-time approximation (RTA) as implemented in the BoltzTraP package~\cite{boltztrap}. The equations for $S$, $\sigma$, and
$\kappa$ are expressed as~\cite{goldsmid,liao,li},
\begin{equation}
  S = -\frac{1}{qT}
  \frac{\sum^{~}_{i, \mathbf{k}}
    (E_{i\mathbf{k}}-\mu) v^{2}_{i \mathbf{k}}
    \frac{\partial f_{i \mathbf{k}}}{\partial E_{i\mathbf{k}}}}
  {\sum^{~}_{i, \mathbf{k}} v^{2}_{i \mathbf{k}}
    \frac{\partial f_{i\mathbf{k}}}{\partial E_{i \mathbf{k}}}}, 
\label{eq:seebeck}
\end{equation}  
\begin{equation}
  \sigma  = -\frac{q^2}{NV} \sum^{~}_{i, \mathbf{k}}
  \tau v^{2}_{i\mathbf{k}} \frac{\partial f_{i \mathbf{k}}}{\partial E_{i\mathbf{k}}},
\label{eq:sigma}
\end{equation}
\begin{equation}
  \kappa_{e}= \frac{1}{NV} \sum^{~}_{i, \mathbf{k}} -
  \frac{\left(E_{i\mathbf{k}}-\mu \right)^{2}}{T} v^{2}_{i\mathbf{k}}
  \tau \frac{\partial f_{i\mathbf{k}}}{\partial E_{i\mathbf{k}}}-TS^{2}\sigma,
\label{eq:kappae}
\end{equation}
where $q = \pm 1.602 \times 10^{-19}~\mathrm{C}$ is the fundamental carrier charge \textbf{of the electron and hole},
respectively, $T$ is the average temperature of the material, $N$ is the
number of $\mathbf{k}$ points in the Brillouin zone, $V$ is the volume
of a unit cell, $\mu$ is the Fermi energy, $f_{i\mathbf{k}}$ is the
Fermi--Dirac distribution, $v_{i\mathbf{k}}$ is the 
electronic group velocity at a particular direction, and $\tau$ is the relaxation-time constant.  We assume that scattering rate
$\tau^{-1}$ is proportional to the temperature.  We should also note that the summations in Eqs.~(\ref{eq:seebeck})--(\ref{eq:kappae}) are performed over not only all possible $\mathbf{k}$ points but also band index $i$.  
 
Phonon thermal conductivity $\kappa_{ph}$ is calculated using the
Debye--Callaway model and RTA.  Its formula is~\cite{kappa-book}
\begin{equation}
  \kappa_{ph} = \frac{k_B}{2\pi^2 \nu}\lp
  \frac{k_B}{\hbar}\rp^3 T^3 \int_0^{\Theta_D/T}
  \tau_{ph}(x) \frac{x^4 e^x}{\lp e^x-1\rp ^2} dx,
\label{eq:kappaph}
\end{equation}
where $x=\hbar\omega/ k_B T$, $k_B$ is the Boltzmann constant, $\nu$
is the effective sound velocity, $\hbar$ is the reduced Plank
constant, $\Theta_D=\hbar\omega_D/k_B$ is the Debye temperature with
$\hbar \omega_D$ being the largest phonon energy of the system, and
$\tau_{ph}$ is the phonon relaxation time.  In this model, we assume
that the Umklapp scattering dominates over the other scattering events, and
it takes the following relaxation-time form~\cite{umklapp}:
\begin{equation}
  \tau_{ph}^{-1}(\omega)=\frac{\hbar\gamma^2}{Mv^2\Theta_D} \omega^2 T,
\end{equation}
where $\gamma$ is the Gr{\"u}neissen parameter and $M$ is the average
atomic mass.

\section{Results and Discussion}
\subsection{CMO geometrical and electronic structure}
\label{sec:geo}

\begin{figure}[t]
  \centering \includegraphics[width=10cm,clip]{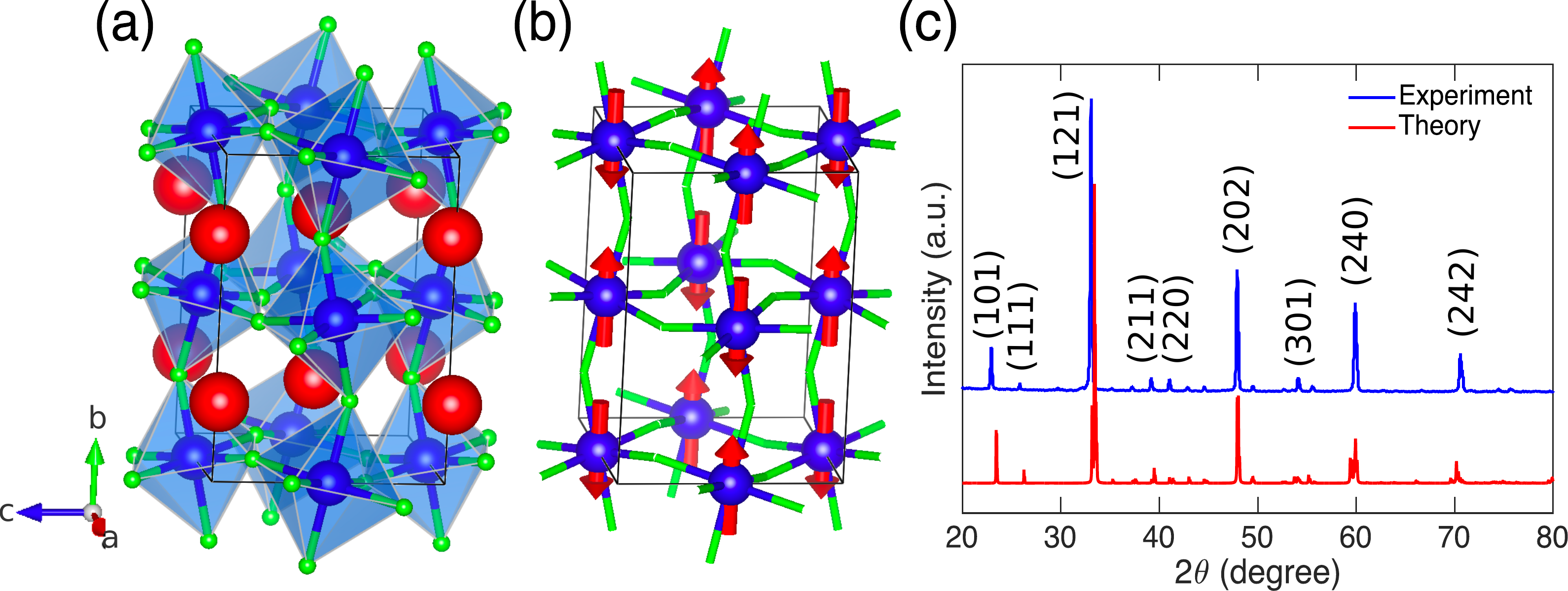}
  \caption{Characterization of the CMO geometrical structure.  (a)
    Geometrical structure of CMO. The (red) Ca atoms 
    are located at the center of the orthorhombic lattice, the (blue) Mn atoms
    are at the edges, and the (green) O atoms surround the Mn atoms, making an
    octahedral structure. (b) CMO exhibits an AFM ground state. 
    (c) XRD pattern of CMO. The (blue line) experimental XRD plot is compared with the (red line) theoretical XRD. The orientation of each peak is given in terms of the Miller index.}
\label{xrd-cmo}
\end{figure}

Figure~\ref{xrd-cmo}(a) shows the orthorhombic perovskite unit
cell of pristine CMO.  In this structure, the Mn ions are surrounded by
six O ions, making an octahedral coordination.  The magnetic
properties of this material originate from the 3d orbitals of the Mn$^{4+}$
ions.  At the ground state, CMO favors AFM ordering because of the spin
configuration of the Mn sublattices~\cite{goodenough}, as shown in
Fig.~\ref{xrd-cmo}(b).  In the present study, we attempt to optimize the
geometrical structure using the total energy minimization method.  The
relaxed structure is compared with the XRD of the CMO sample shown in
Fig.~\ref{xrd-cmo}(c).  The blue line shows the experimental results,
whereas the red line is obtained from DFT, which are in good agreement
with those in the experiment.  The calculated lattice constants are
$a=5.3265 \unitangstrom$, $b=7.4319 \unitangstrom$, and
$c=5.2523 \unitangstrom$, with a discrepancy of approximately $0.5 \%$ from the
experimental results.

\begin{figure}[t]
  \centering \includegraphics[width=7cm,clip]{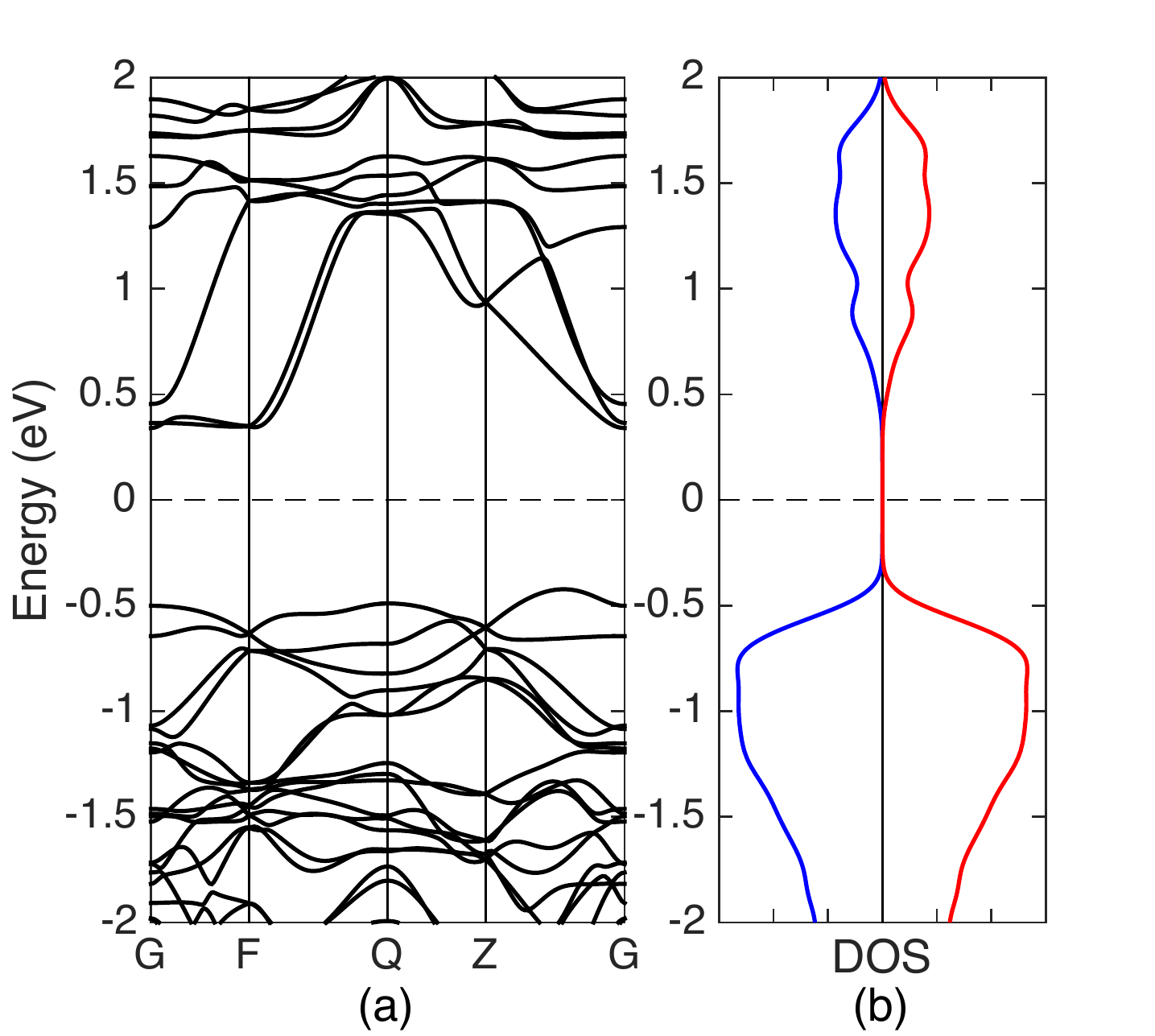}
  \caption{Electronic structure of CMO. (a) Energy dispersion.  (b)
    Density of states. The blue and red lines refer to the spin-up and spin-down
    states, respectively.}
\label{dos-cmo}
\end{figure}

The electronic structure is calculated by considering the relaxed
structure.  We obtain the electronic-band structure of CMO as shown in
Fig.~\ref{dos-cmo}(a), which is consistent with the previous
calculations~\cite{zhang2011}. \textbf{From the result, we obtain an indirect bandgap of approximately $0.76 \unitev$.
Compared with our temperature of interest in the range of $300$--$500$ K, the bandgap is relatively small.
At the ground state, CMO exhibit an AFM insulating state, which is in good agreement with
previous optical spectroscopy results~\cite{jung1997}. 
The DOS of CMO in the spin-up (blue line) and spin-down (red line) cases are shown in
Fig.~\ref{dos-cmo}(b).}

\subsection{CMO electronic transport}
\label{sec:el1}
\begin{figure}[t]
\centering \includegraphics[width=10cm,clip]{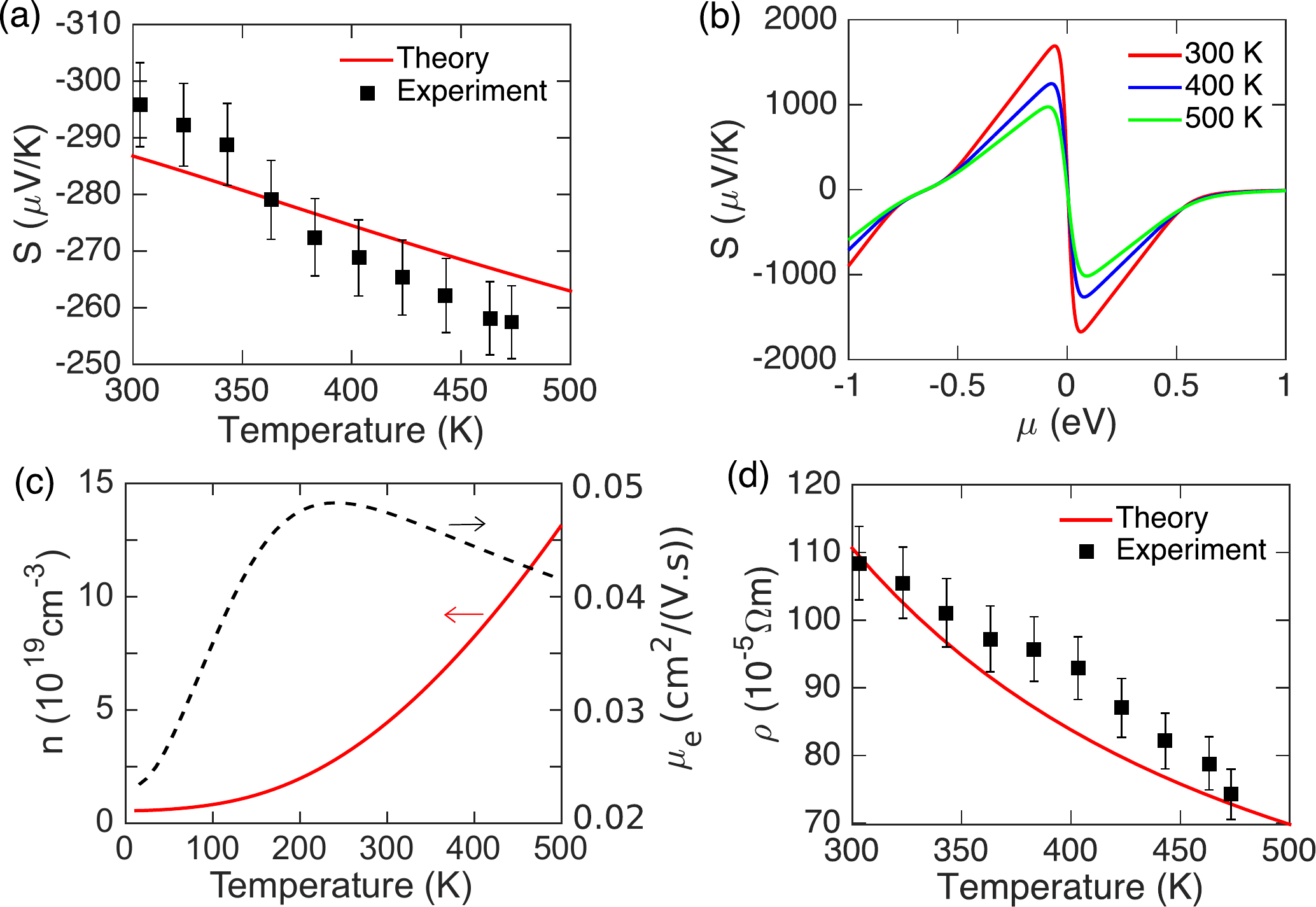}
\caption{Transport properties of pristine CMO.  (a) The (dots) experimental
  Seebeck coefficients $S$ are compared with the (red line) theoretical
  results at chemical potential $\mu=0.5\unitev$. The
  $y$-axis is inverted to emphasize the absolute value of $S$ while
  keeping the negative sign for the $n$-type semiconductors.  (b) Plot of
  theoretical $S$ as a function of chemical potential $\mu$ under
  RTA.  At small $\mu$, the Seebeck
  coefficient can be optimized to reach $\sim 1600\ \mu {\rm V/K}$.
  (c) Plot of the (solid red line) carrier concentration that optimizes $S$  
  and (black dashed line) carrier mobility as a
  function of temperature.  (d) (Dots) experimental resistivity $\rho$ 
  compared with (red line) theoretical resistivity in
  Eq.~(\ref{eq:sigma}). We have adopted $\tau^{-1}\propto T$ in the
  theoretical model. }
\label{seebeck-cmo}
\end{figure}

From the electronic-band structure, we can obtain the charge concentration and group velocity of the electrons to determine the electronic transport properties of the materials, according to Eqs. (\ref{eq:seebeck})--(\ref{eq:kappae}). The calculated Seebeck coefficients obtained from Eq.~(\ref{eq:seebeck}) are shown in Fig.~\ref{seebeck-cmo}(a) and compared with the experimental results (dots). The Seebeck coefficients modeled in Eq.~(\ref{eq:seebeck}) only depend on Fermi energy $\mu$ and the temperature.  By fitting this coefficients with the experimental results of $S$ versus $T$, we obtain Fermi energy $\mu=0.5\unitev$, which is  measured from the theoretical charge neutrality point (horizontal dashed line) shown in Fig.~\ref{dos-cmo}(a).  This result means that the Fermi energy is at the edge of the conduction band. Although we obtain $\mu$ from a self-consistent DFT calculation, shifting $\mu$ within the gap does not change the geometrical structure at the ground state.

The negative sign of the Seebeck coefficient indicates
that CMO is $n$-doped.  The inverse proportionality of $|S|$ with respect
to $T$ originates from the semiconducting or insulating nature of
CMO~\cite{goldsmid}.  This inverse proportionality also persists in
different $\mu$.  The sign change of $S$ occurs due to the ambipolar
effect following the electron or hole charge as majority or minority
carriers~\cite{nguyen2015}.

Theoretically, the Seebeck coefficient may be amplified by one order
of magnitude if the Fermi energy is located near $\mu=0$, as shown in
Fig.~\ref{seebeck-cmo}(b).  Such a high Seebeck coefficient has been
obtained in organic perovskites and can be achieved
in perovskite oxides~\cite{ye2017}.  Adjusting the Fermi energy to
optimize $|S|$ involves modifying the carrier concentration or doping the
system.  In CMO, we find that the corresponding carrier concentration
at room temperature should be approximately $5 \times 10^{19}$ cm$^{-3}$.
Furthermore, the carrier mobility ($\mu_e$) can be directly derived from the formula $\mu_e = \sigma / n e$, where $\sigma$ is the electron conductivity, $e$ is the absolute value of the electron charge, and $n$ is the carrier concentration. The carrier concentration and carrier mobility as a function of temperature are plotted and shown in Fig.~\ref{seebeck-cmo}(c), which shows that the mobility values in CMO vary within $0.02$--$0.05$ $\mathrm{cm}^2/\mathrm{V.s}$ and reach an optimum value at a temperature near $200$ K.  

We can further determine electrical resistivity
$\rho = 1/\sigma$ from Eq.~(\ref{eq:sigma}) and compare the
calculated result with the experimental result.  
Figure~\ref{seebeck-cmo}(d) shows $\rho$ as a function of temperature.
We obtain $\tau^{-1}=\alpha T + \beta$ with $\alpha = 0.008$ ${\rm
  fs}^{-1}{\rm K}^{-1}$ and $\beta=0.4$ ${\rm fs}^{-1}$. Transport
lifetime $\tau$ is proportional to the transport mobility and is needed to
characterize the number of defects and device performance. $\tau^{-1}$ is proportional to the temperature, as expected, in the electron--phonon scattering, i.e., the number of phonons increases due to the thermal excitation.  The decrease in $\rho$ as a function of temperature is expected from the
  Drude model of semiconductors. $\rho$ is inversely proportional to the carrier concentration, whereas the carrier concentration increases due to the thermal excitation.

\begin{figure}[t]
    \centering
    \includegraphics[width=9cm,clip]{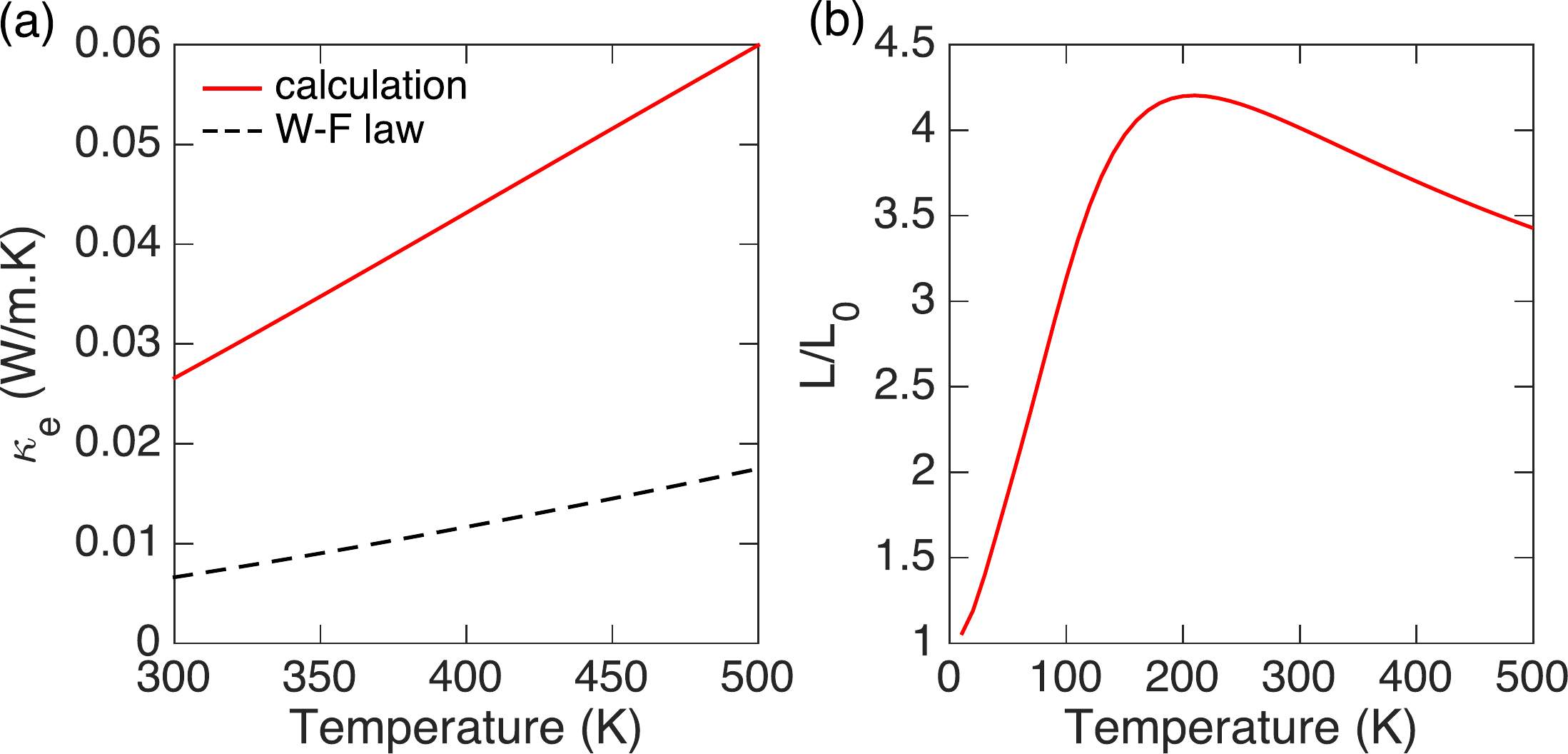}
    \caption{(a) (solid red line) Electronic thermal conductivity $\kappa_e$ 
      of pristine CMO as a function of temperature obtained from
      Eq.~(\ref{eq:kappae}) compared with the (black dashed line)
      prediction from the WF law. (b)
      Calculated Lorenz number as a function of temperature in units
      of $L_0=(\pi k_B/e)^2/3$.}
    \label{kappa_cmo}
\end{figure}

According to the previously obtained parameters ($\mu$ and $\tau$), we show calculated electronic thermal conductivity $\kappa_e$ from
Eq.~(\ref{eq:kappae}) in Fig.~\ref{kappa_cmo}(a).  We can compare the
electron efficacy in the conducting heat versus current using the WF law, i.e., 
$\kappa_e = L T/\rho$, where $L$ is the Lorenz number.  For ordinary
metals, we have $\kappa_e^{(0)}=L_0 T/\rho$, where
$L_0=(\pi k_B/e)^2/3$.  Introducing $\rho$ to that shown in
Fig.~\ref{seebeck-cmo}(d), we obtain $\kappa_e^{(0)}$ as indicated by the dashed
line shown in Fig.~\ref{kappa_cmo}(a).  The $\kappa_e$ value is typically
approximately four times larger than expected $\kappa_e^{(0)}$ from the WF
law with temperature dependence, as shown in Fig.~\ref{kappa_cmo}(b).
These results imply that the localized electrons in the Mn orbitals
more efficiently conduct heat than the current, resulting in a
drawback for TE application.  Violation of the WF law due to strong correlation~\cite{mahajan2013,principi2015,lavasani2019} and bipolar effects~\cite{yoshino} has been
reported. However, the bipolar effects show weak dependence on $L/L_0$ versus $T$. Thus, we attribute the breakdown of the WF law to the strong correlation.  The relationship of $L$
with the strength of Coulomb interaction $U$ is very instructive to
study in detail. However, to understand this phenomenon, an in-depth
investigation will be presented elsewhere because it is not of primary importance in the present study where the $\kappa_e$ value is negligible compared with
the total $\kappa$ value shown in Fig.~\ref{ZT}.  As shown in
Fig.~\ref{kappa_bcmo}, the Bi substitution at the Ca site reduces the $L/L_0$
ratio, hence the recovery of the metallic nature.

\subsection{BCMO electronic structure and transport}
\label{sec:el2}

\begin{figure}[t]
  \centering \includegraphics[width=10cm,clip]{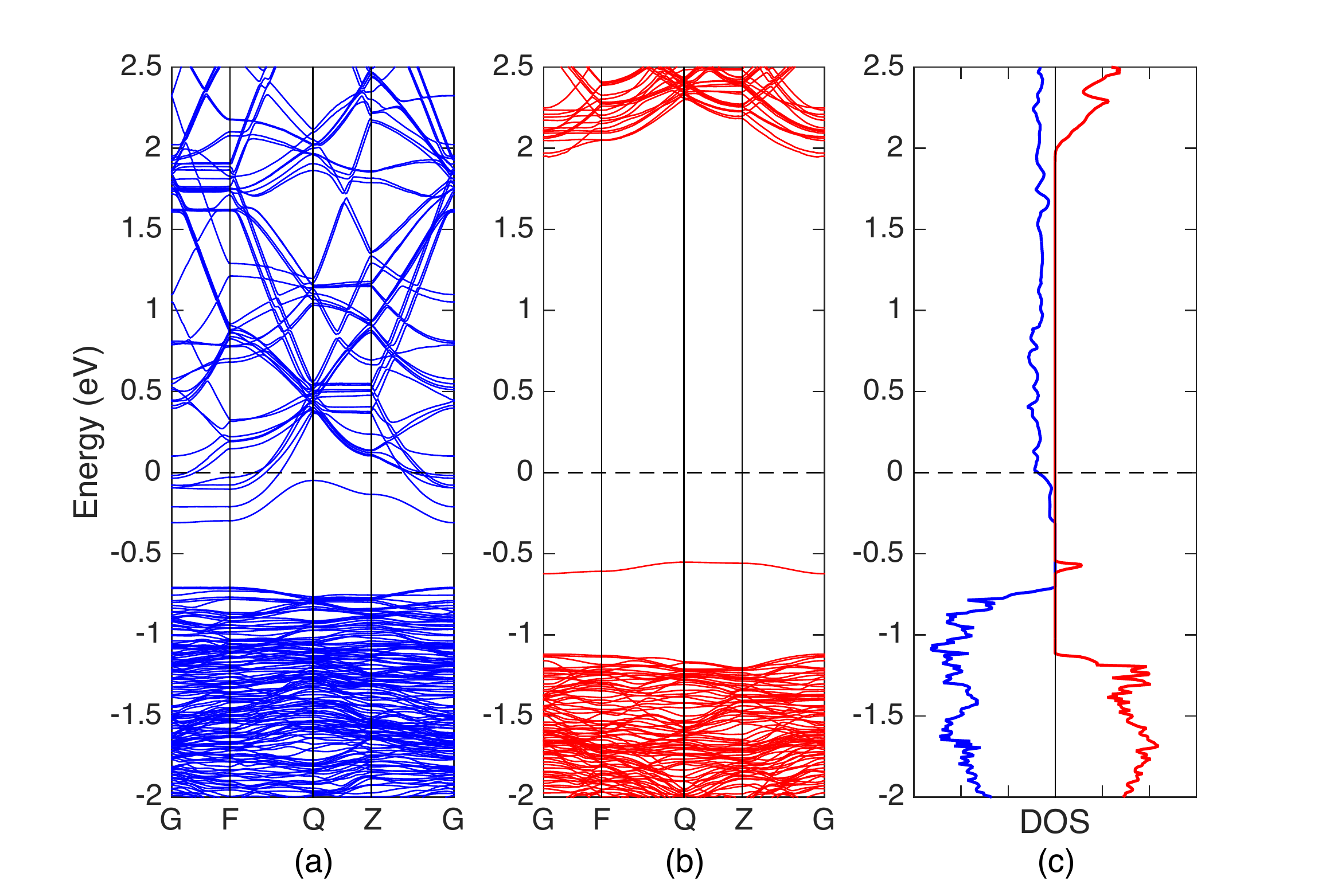}
  \caption{Electronic band structure of BCMO for (a) spin-up and (b)
    spin-down states. (c) Density of states of BCMO. The blue and red
    lines refer to the spin-up and spin-down states, respectively.}
\label{dos-bcmo}
\end{figure}

One way of improving the TE properties of CMO is by performing
atomic substitution of Bi at the Ca site.  With this substitution, the
two different ions, namely, $\rm Mn^{3+}$ that sits next to the Bi ion and
$\rm Mn^{4+}$ near the Ca ions, perform a double
exchange (DE) mechanism mediated by
oxygen~\cite{Loshkareva2012}. This DE interaction delocalizes the
electrons in the Mn $d$ orbitals and favors the metal ferromagnetic ground
state~\cite{anderson}, as shown in the DOS of
BCMO in Fig.~\ref{dos-bcmo}(c). From the electronic-band structure
results shown in Figs.~\ref{dos-bcmo}(a) and (b), the Fermi energy lies in
the conduction band with an imbalanced spin-up population over the
spin-down component. As a result, the electronic transport properties
are expected to mimic a metallic Fermi gas. Nevertheless, the
strongly correlated nature remains, as shown in the reduced band
gap ($\sim 0.5\unitev$) below the Fermi level because of on-site Coulomb
interaction $U$.

\begin{figure}[t]
  \centering\includegraphics[width=10cm,clip]{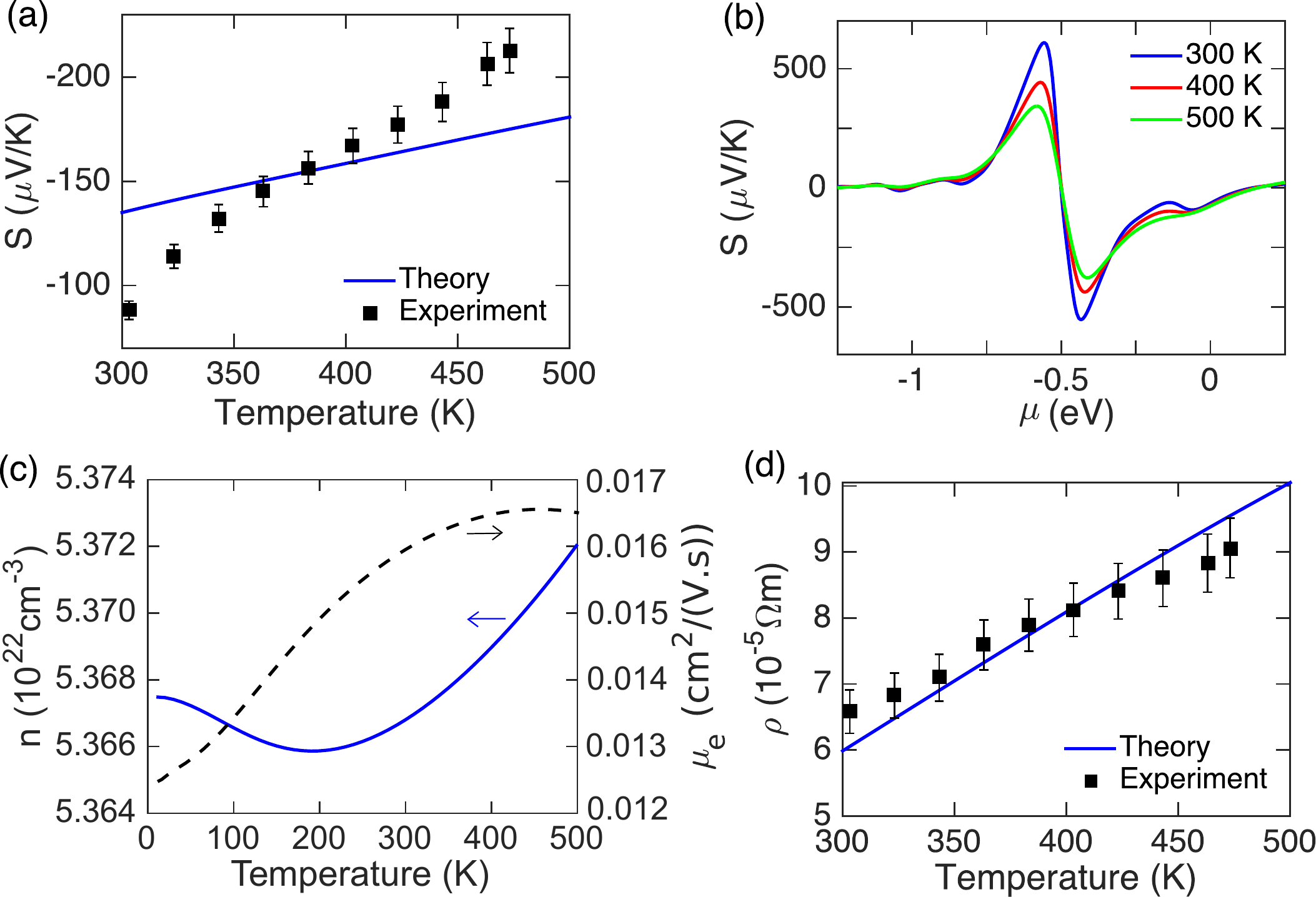}
  \caption{Transport properties of pristine BCMO.  (a) (blue line) Seebeck
    coefficient $S$ as a function of temperature from
    Eq.~(\ref{eq:seebeck}) and (dots) the experimental results.
    (b) Plot of theoretical $S$ as a function of chemical
    potential $\mu$. (c) Plot of (solid blue line) carrier concentration and 
    (black dashed line) carrier mobility as a function
    of temperature.  (d) (Dots) experimental $\rho$ compared with
    (blue line) theoretical $\rho$ in Eq.~(\ref{eq:sigma}).}
    \label{seebeck-bcmo}
\end{figure}

In contrast to that of pristine CMO, the Seebeck coefficient of BCMO
is proportional to the temperature [see Figs.~\ref{seebeck-bcmo}(a) and 
\ref{seebeck-cmo}(a)], as expected from the linear scaling in metal
$|S|=\pi^2 k_B T/2e \mu$~\cite{goldsmid}.  In this case, the Seebeck
coefficient does not depend on $\tau$. Thus, we can fit $S$ with the
experimental data to obtain $\mu$.  We obtain $\mu=-0.25 \unitev$ 
from the charge neutrality point, which means that
the Fermi energy is located at the edge of the conduction band. The
optimum Seebeck coefficient is approximately $500\unitseebeck$ or approximately four
times smaller than that of the pristine sample. This optimum value can
be achieved by positioning the Fermi energy near $\mu=-0.5\unitev$ at the
band gap and low temperature, as shown in
Fig.~\ref{seebeck-bcmo}(b). At room temperature, the carrier
concentration of BCMO is approximately $5 \times 10^{22}$ cm$^{-3}$, as shown
in Fig.~\ref{seebeck-bcmo}(c). It has $10^{3}$ times more
concentration than pristine CMO because of the Bi doping.
We also find that the mobility values of BCMO vary within $0.012$--$0.017$ $\mathrm{cm}^2/\mathrm{V.s}$ and tend to increase with increasing temperature until reaching a maximum value at a temperature between $400$ and $500$~K, as shown in Fig.~\ref{seebeck-bcmo}(c).

Opposite behavior of CMO with respect to BCMO transport can also be
observed from electrical resistivity $\rho$ profile that proportionally increases 
with $T$, as shown in Fig.~\ref{seebeck-bcmo}(d). For a conventional metal with
$\mu\gg k_BT$, the electrical conductivity takes the form of 
$\sigma =\rho^{-1}=\frac{e^2}{2\pi \hbar^2} \tau \mu$. Thus, the only
dependence of $\rho$ on $T$ comes from the temperature dependence of
relaxation time ($\tau$). Indeed, by considering the relaxation rate as
$\tau^{-1}=\alpha T+\beta$, 
calculated $\rho$ reproduces the experimental results with $\alpha=0.013$
${\rm fs}^{-1}{\rm K}^{-1}$ and $\beta=0.2$ ${\rm fs}^{-1}$. We parenthetically note
that the scattering rate of BCMO is 1.5 times larger than
that of CMO. In other words, the electron lifetime in BCMO is shown to be reduced 
due to the Bi doping.  Notably, despite the reduction in the electronic
lifetime, the $\rho$ values of BCMO decrease by one order of magnitude
compared with the CMO because of the metallic nature of BCMO. Hence,
better TE performance is achieved.

\begin{figure}[t]
  \centering \includegraphics[width=9cm,clip]{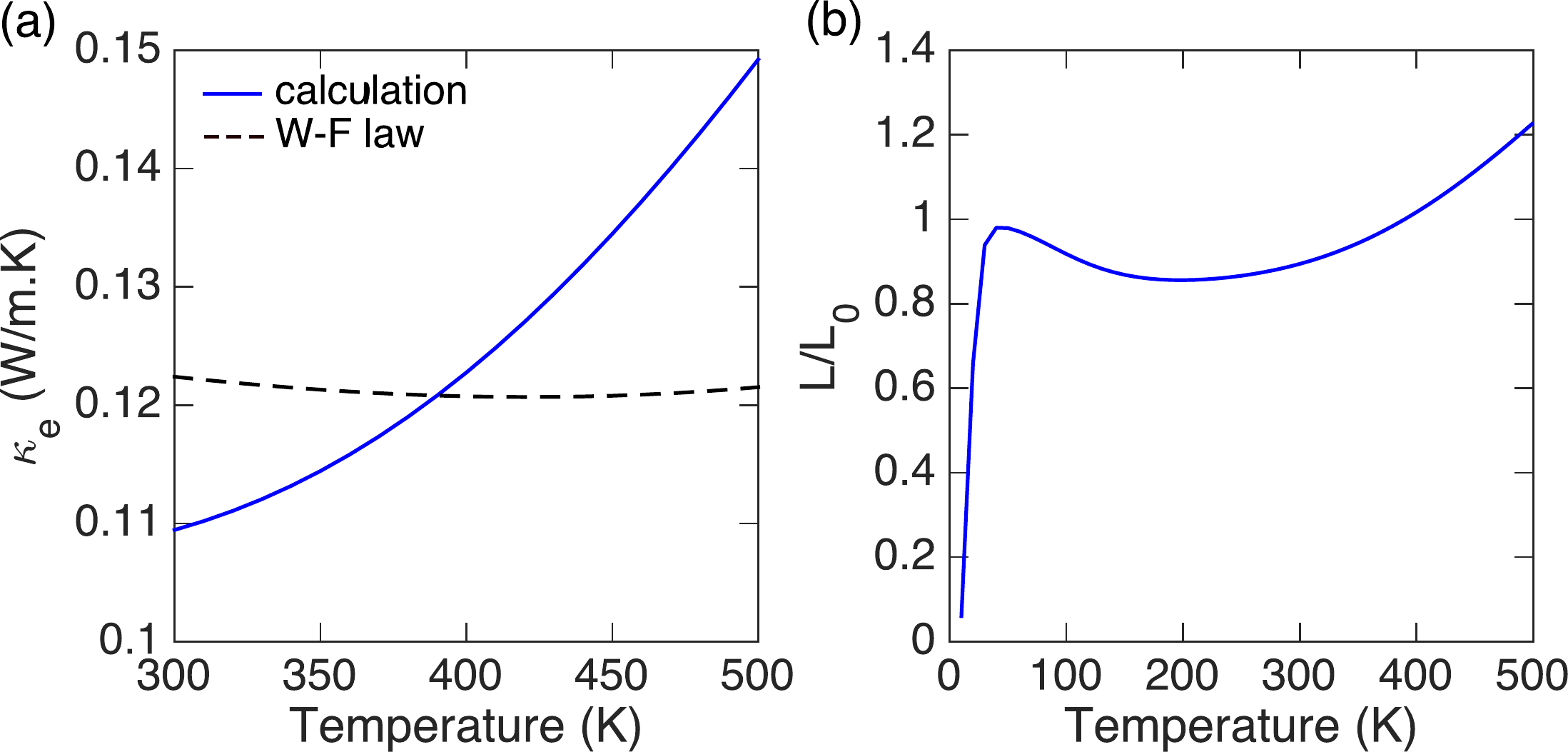}
  \caption{(a) (Blue solid line) electronic thermal conductivity $\kappa_e$ of BCMO as a
    function of temperature from Eq.~(\ref{eq:kappae}) 
    compared with the (black dashed line) prediction from the WF law.
    (b) Calculated Lorenz number as a function of
    temperature compared with $L_0=(\pi k_B/e)^2/3$. }
    \label{kappa_bcmo}
\end{figure}

The electronic thermal conductivity of BCMO is shown to be linearly
proportional to the temperature. However, this trend cannot be reproduced
from the WF law [see Fig.~\ref{kappa_bcmo}(a)] because this system
also possesses a Lorenz number that depends on the temperature
[Fig.~\ref{kappa_bcmo}(b)]. Nevertheless, the $L/L_0$ ratio does not 
strongly depart from unity, which signifies the contribution from the Bi
doping to recover the metallic characteristic of CMO. In contrast to
pristine CMO, $L/L_0$ can be less than unity at a temperature below 400 K.

\subsection{Phonon contribution to thermal conductivity}
\label{sec:ph}

\begin{figure}[t]
  \centering
  \includegraphics[width=10cm,clip]{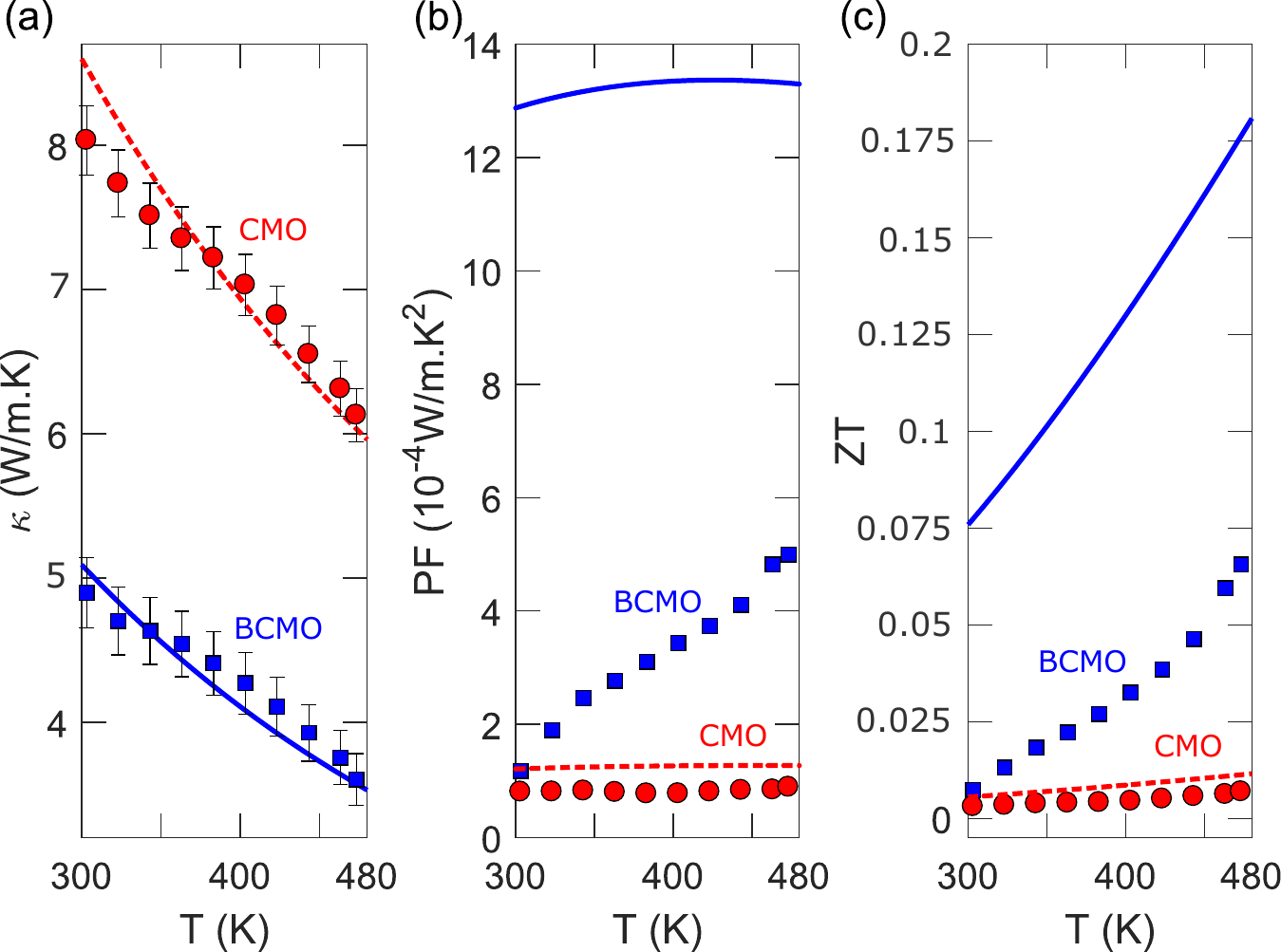}
  \caption{TE properties of CMO and BCMO. The experimental
    (theoretical) results are denoted by symbols (lines). The red circles
    and dashed lines refer to CMO, whereas blue squares and solid lines
    refer to BCMO. Panel (a) shows experimental thermal conductivity
    $\kappa$ and theoretical phonon thermal conductivity $\kappa_{ph}$
    in Eq.~(\ref{eq:kappaph}).  The electronic thermal conductivity can
    be neglected in the fitting because its values are approximately two
    orders of magnitude smaller than $\kappa_{ph}$. In panels (b) and
    (c), the measured power factor and dimensionless
    figure of merit, respectively, are compared with the optimized values 
    (or maxima) that can be achieved by the
    theoretical prediction.}
    \label{ZT}
\end{figure}

Using the simplified Callaway--Debye model in Eq.~\ref{eq:kappaph} and
assuming that the Umklapp scattering is dominant, we can reproduce the
measured thermal conductivity of pristine CMO, as shown in
Fig.~\ref{ZT}(a).  The Debye frequency is assumed to be equal to the largest
phonon peak in the Raman spectra (see \ref{sec:raman}).  For
pristine CMO, we use the Debye temperature of $800\ \rm K$ and Debye
frequency of $557\ \rm cm^{-1}$ from the Raman spectra.  We also assume 
Gr{\"u}neissen parameter $\gamma=3.45$ from a previous work
\cite{srivastava2009}.  We then obtain effective sound velocity
$\nu=1.38\ \rm km/s$, which is typically comparable with a similar
work by an Israel group~\cite{baranovskiy}.  For BCMO, we use the Debye
temperature of 790 K (corresponding to the red-shifted Raman peak of 
$549\ \rm cm^{-1}$) and Gr{\"u}neissen parameter $\gamma=3.47$.  The
sound velocity thus obtained for BCMO is $\nu=0.81\ \rm km/s$.

We must note that Bi doping decreases thermal conductivity $\kappa$
and electrical resistivity $\rho$.  The decrease in $\rho$ by $10$ times
can overcome the Seebeck coefficient by two times, which
makes BCMO power factor $\mathrm{PF}=S^2/\rho$ to be superior compared with
pristine CMO, as shown in Fig.~\ref{ZT}(b). We show in this figure, that the
power factor of BCMO may, in principle, be amplified up to two times
compared with that in the experiment if the Fermi level is located at the band
gap with $\mu=-0.5\unitev$ from the charge neutrality, as shown in
Fig~\ref{seebeck-bcmo}.  The dimensionless figure of merit,
$ZT=S^2T/\rho\kappa$ in Fig.~\ref{ZT}(c) shows that the TE 
performance in CMO can be improved by Bi doping. 

Figures.~\ref{ZT}(b) and (c) show that our current experiments have not yet reached the optimized values of PF and ZT from the theoretical prediction.  We roughly estimate that to approach the optimized (theoretical) $\mathrm{PF}$ and $ZT$, a subfraction of doping within $2.8\%$--$3.2\%$ might be needed in the experiment. However, in reality, achieving the optimum doping by atomic substitution is highly complicated. The presence of Bi doping facilitates the DE mechanism. The Bi doping dramatically alters the CMO phase from an AFM insulator to an FM conductor. A slight change in the doping concentration might result in an abrupt change in the electronic-band structure. Thus, tuning the optimum doping concentration according to the theory could be difficult. Moreover, the measurement of carrier concentration can be performed using the Hall effect setup. However, this technique is only applicable to thin films and not to bulk samples. Nevertheless, we expect that the theoretical prediction of optimized PF and $ZT$ can trigger further experiments in the future to adjust the doping level (or chemical potential or Fermi energy) so that obtaining  optimized (better) values of PF and $ZT$ becomes possible.

Our TE measurements show that CMO has a reasonably high Seebeck
coefficient of approximately $300$~${\rm \mu V/K}$ at room temperature, and 
optimizing the Seebeck coefficient up to five times is even possible 
if the chemical potential is correctly tuned (at the band edge). By
Bi doping, \textbf{the electrical resistivity is reduced by an order of magnitude}, and the
Seebeck coefficient becomes two times smaller than thatin pristine CMO, hence significantly 
improving the power factor.  Additionally, the Bi
doping adds more phonon scattering path via anharmonicity and
increases the atomic average mass.  As a result, the phonon
  lifetimes and thermal conductivity are reduced, as illustrated in the
  transport measurements. Combined with the reduced electrical
  resistivity, the TE figure of merit of BCMO is enhanced.  Another
notable phenomenon is the breakdown of WF law in CMO, where Lorenz
number $L_0$ reaches four times that of conventional metals. The breakdown
of the WF law comes from the strongly correlated nature of electrons in CMO,  
\textbf{which suggests that the strong electron correlation in CMO separates the heat and charge transport.}
Moreover, the particular shape of the bands and scattering profiles can significantly differ from the metallic Lorenz value ~\cite{wang2018}.  During the Bi-doping, this Lorenz number recovers from that of ordinary metals.

\section{Conclusions} 
\label{sec:conclusion}
We have investigated the electron and phonon transport properties of $\cmo$ and $\bcmo$ using TE transport measurements and first-principles calculations. Our calculations indicate that $\cmo$ has AFM insulating ground states, which results in a significant Seebeck coefficient and electrical resistivity. The strongly correlated nature of $\cmo$  manifests in the breakdown of the WF law in which the Lorenz number is four times larger than that of ordinary metals. In the Bi-doping, the electrical resistivity and conductivity decreases and increases, respectively, by an order of magnitude, whereas the Seebeck coefficient decreases by two times, providing a more prominent power factor than the pristine one. $\bcmo$ also recovers the correspondence with the WF law. The presence of Bi doping decreases the sound velocity as well as the phonon lifetime. Overall, the Bi doping enhances the TE figure of merit because of the improved electronic transport and phonon thermal-transport degradation.

\section*{Acknowledgments}
This research was partially funded by a collaboration grant from the
secretariat of Indonesian Institute of Sciences (LIPI) for a visit of
the LIPI researchers to Thailand.  We acknowledge HPC LIPI for computational facility. M.~N. acknowledges a research grant
from Brawijaya University No. 3/UN10.F09/PN/2019. B.~K. acknowledges
Ministry of Research and Technology/National Research and Innovation Agency of Indonesia (Kemenristek
/ BRIN) with PDUPT 2020 grant [NKB-2799/UN2.RST/HKP.05.00/2020].
W.~B.~K.~P. acknowledges Raman research facilities in LIPI Research
Center for Physics.  T.~S. and K.~S. acknowledge TRF Research Career
Development Grant RSA6180070.

\appendix
\section{Raman Spectroscopy of CMO and BCMO}
\label{sec:raman}

\begin{figure}
  \centering
  \includegraphics[width=68mm,clip]{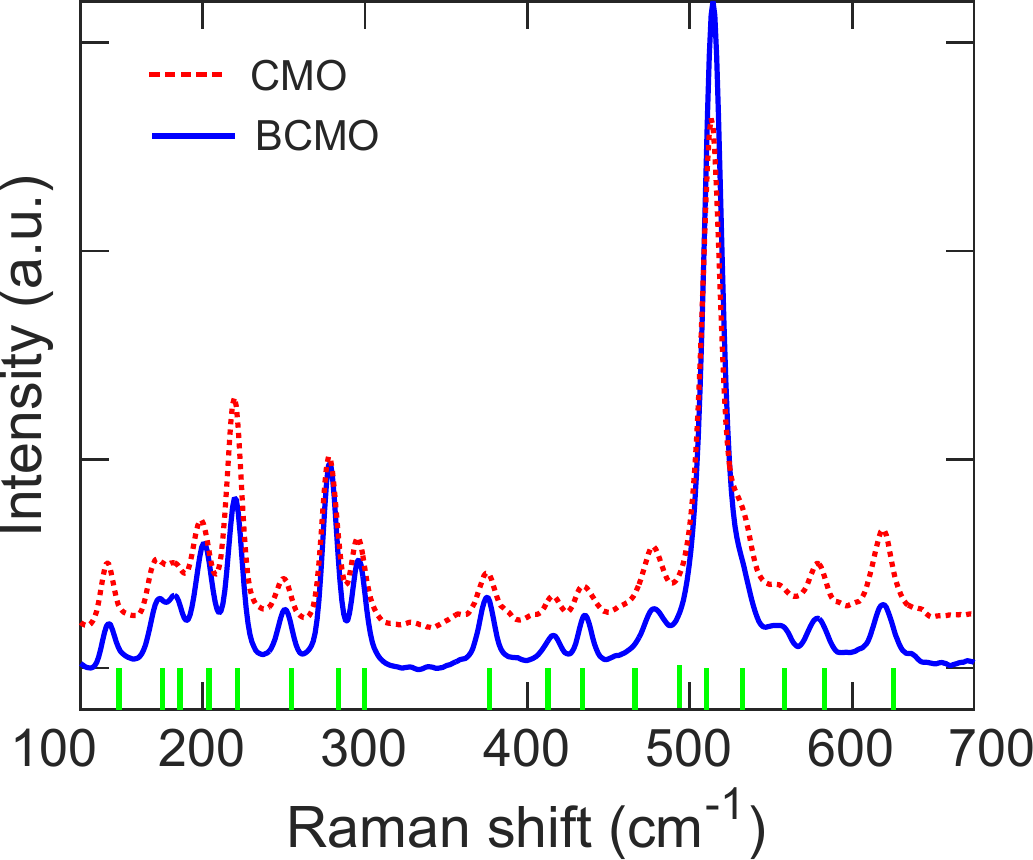}
  \caption{Experimental Raman spectroscopy of pristine CMO (red
   dash line) and BCMO (blue solid line). Short vertical green lines indicate
    the frequencies of theoretically Raman active modes in CMO.}
  \label{raman}
 \end{figure}

In addition to the TE measurements, we experimentally studied the phonon properties of CMO and BCMO via Raman spectroscopy. CMO and BCMO were prepared according to the standard sol--gel method. First, we mixed the stoichiometric reagents of $\textrm{Ca}(\textrm{NO}_3)_2 \cdot 4 \textrm{H}_2\textrm{O}$ (Merck, purity $\geq 99 \%$), $\textrm{Mn}(\textrm{NO}_3)_2 \cdot 4 \textrm{H}_2\textrm{O}$ (Merck, purity
$\geq 98.5 \%$), Citric acid $\textrm{C}_6\textrm{H}_8\textrm{O}_7 \cdot \textrm{H}_2\textrm{O}$ (Merck, purity $\geq 99.5 \%$), $\textrm{Bi}_2\textrm{O}_3$ (Fluka, purity $\geq 99.8 \%$), and distilled water.  The dried gel was crushed and ground into a powder and then calcined at $923$ K for $6$ h in air.  Then, the powder was die-pressed into square-shaped pellets and sintered at $1273$ K.  To ensure quality
of the final product, another round of sintering was applied at $1473$ K
for $12$ h, which were completed in the air atmosphere. Raman
spectra were collected at room temperature using the iHR 320 modular Raman (Horiba Jobin Yvon) spectrometer equipped with a CCD detector and a solid-state laser source that operated at 532 nm wavelength with
appropriate filters. The spectrometer had a spectral resolution of
$1.22$ $\textrm{cm}^{-1}$/pixel, 600 gr/mm grating, and an objective of 100$\times$.

Figure~\ref{raman} shows the experimental Raman spectra of CMO and BCMO of the unpolarized light. 
The short lines below the spectra are the theoretical Raman active modes. 
Our results qualitatively reproduced those of Ref.~\cite{abrashev}.
We also observed a significant increase in the $513\ \rm cm^{-1}$ mode intensity from the Bi-doping.
This peak is was due to defect scattering~\cite{abrashev}. Therefore, we can expect an increase in intensity of BCMO. For the details of the modes, Table \ref{Tab1} lists the complete Raman active modes in CMO and BCMO along with their frequencies.

\begin{table}
\caption{Raman active modes of CMO and BCMO.
\label{Tab1}}
\begin{center}
\begin{tabular}{cccc}
\hline

\multirow{2}{*}{symmetry} &
\multicolumn{3}{c}{frequencies ($\mathrm{cm}^{-1}$)} \\
{} & theory & exp. CMO & exp. BCMO\\
\hline
A$_g$    & 148 & 141 & 142\\
B$_{2g}$ & 176 & 171 & 172\\
B$_{1g}$ & 186 & 183 & 184\\
B$_{3g}$ & 203 & 199 & 201\\
B$_{2g}$ & 221 & 220 & 221\\
A$_g$    & 254 & 250 & 251\\
A$_g$    & 284 & 278 & 279\\
A$_g$    & 300 & 295 & 296\\
B$_{2g}$ & 376 & 376 & 375\\
B$_{2g}$ & 413 & 416 & 417\\
B$_{3g}$ & 434 & 434 & 435\\
B$_{2g}$ & 466 & 476 & 477\\
B$_{1g}$ & 494 & 500 & 499\\
A$_g$    & 506 & 513 & 514\\
A$_g$    & 532 & 533 & 531\\
B$_{1g}$ & 558 & 556 & 558\\
B$_{1g}$ & 582 & 580 & 579\\
B$_{2g}$ & 625 & 620 & 619\\
\hline
\end{tabular}
\end{center}
\end{table}

%\bibliography{mybibfile}

\end{document}